# Northwest Africa 5958: a weakly altered CM-related ungrouped chondrite, not a CI3

Emmanuel Jacquet[1,2], Jean-Alix Barrat[3], Pierre Beck[4], Florent Caste[1], Jérôme Gattacceca[5], Corinne Sonzogni[5], Matthieu Gounelle[1,6].

[1]Institut de Minéralogie, de Physique des Matériaux et de Cosmochimie, CNRS & Muséum National d'Histoire Naturelle, UMR 7202, 57 rue Cuvier, 75005 Paris, France.
[2]Canadian Institute for Theoretical Astrophysics, 60 St George Street, Toronto, ON, M5S 3H8, Canada.

[3]Université Européenne de Bretagne, France; CNRS UMR 6538 (Domaines Océaniques), U.B.O.-I.U.E.M., Place Nicolas Copernic, 29280 Plouzané Cedex, France.

[4]1 Univ. Grenoble Alpes, Institut de Planétologie et d'Astrophysique de Grenoble (IPAG), F-38000 Grenoble, France ;CNRS, IPAG, F-38000 Grenoble, France.

[5]CEREGE UM 34, CNRS/Université d'Aix-Marseille 3, 13545 Aix-en-Provence, France.

[6]Institut Universitaire de France, 103 Boulevard Saint Michel, 75005 Paris, France.

E-mail: emjacquet@mnhn.fr

## *Abstract*

Northwest Africa (NWA) 5958 is a carbonaceous chondrite found in Morocco in 2009. Preliminary chemical and isotopic data leading to its initial classification as C3.0 ungrouped have prompted us to conduct a multi-technique study of this meteorite and present a general description here. The petrography and chemistry of NWA 5958 is most similar to a CM chondrite, with a low degree of aqueous alteration, apparently under oxidizing conditions, and evidence of a second, limited alteration episode manifested by alteration fronts. The oxygen isotopic composition, with $\Delta'^{17}O$ = -4.3 ‰, is more $^{16}O$-rich than all CM chondrites, indicating, along with other compositional arguments, a separate parent body of origin. We suggest that NWA 5958 be reclassified as an ungrouped carbonaceous chondrite related to the CM group.

## *1. Introduction*

While chondrites generally represent the original condensable matter of the solar system, they are not all "primitive" to the same degree, or in the same way. Chemically, the chondrites closest to solar photospheric abundances are the carbonaceous chondrites (Palme & Jones 2005), and they are also closer to the Sun in oxygen isotopic composition than others (McKeegan et al. 2011). However, many of them have undergone aqueous alteration (McSween 1977) and/or thermal metamorphism blurring their nebular properties, and ironically, the major chemical group closest to photospheric abundances, namely the CI group, ranks as the most aqueously altered one, with few pre-accretionary components preserved (Gounelle & Zolensky 2014). A real benchmark to understand nebular fractionation processes and thence protoplanetary disk physics is still missing.

Beside the established major groups, considerable interest is thus laid on ungrouped carbonaceous chondrites which may approach this ideal, taking advantage of the surge in meteorite finds from hot and cold deserts (e.g. Choe et al. 2010; Zhang & Yurimoto 2013; Kimura et al. 2014). A notable example is Acfer 094 (e.g. Newton et al. 1995), a CO/CM-like chondrite with very low degree of aqueous alteration or thermal metamorphism (type 3.0). Another is Tagish Lake, a witnessed fall, with CI/CM affinities but petrographic type 2 (Zolensky et al. 2002; Friedrich et al. 2002).

Northwest Africa (NWA) 5958 may be a further milestone in this quest. This single, fragmented fresh stone weighing 286 g discovered in Morocco in 2009 was found to be weakly altered and classified as a C3.0-UNG in the 99$^{th}$ Meteoritical Bulletin (see Bunch et al. 2011). In fact, preliminary ICP-MS analyses suggested it had a CI bulk composition (Ash et al. 2011), despite the presence of chondrules, and the most $^{16}O$-rich composition of all known bulk meteorites (Ash et al. 2011). It is hardly necessary to stress how profound the implications of the identification of such a "CI3", if confirmed, on the relationship between chondrite components would be. In particular, it would challenge "two-component" pictures explaining chondrite compositions as a mixing between a CI chondritic matrix and a high-temperature volatile-depleted component (see e.g. Anders 1964; Zanda et al. 2006; Zanda et al. 2012) and favour matrix/chondrule complementarity scenarios (see e.g. Wasson & Chou 1974; Hezel & Palme 2010; Palme et al. 2015).

Yet, since its original classification in 2010, no study on NWA 5958 has been published except in abstract form (e.g. Bunch et al. 2011; Ash et al. 2011; Nittler et al. 2012; Elmaleh et al. 2012; Stroud et al. 2014; Sanborn et al. 2015), save for a Cr isotopic measurement as part of the study of Göpel et al. (2015). We have thus set out to present our own description of NWA 5958, following a multi-technique study, including, given their potential importance, replications of the chemical and isotopic measurements. Since some of the original results have to be revised in light of this study, we will then discuss anew the classification of this intriguing meteorite.

## 2. Samples and analytical procedures

### 2.1 Sample and petrographic examination

Two specimens of NWA 5958, weighing 4.706 and 4.626 g, were purchased from the Hupé collection and were the source of the subsamples used in this study. Among the polished sections made from them, three, named NWA 5958-1, NWA 5958-5 and NWA 5958-6, were examined specifically in this work. Object names (such as N1-10) start with N1, N5 or N6 depending on their host section. The sections were examined in scanning electron microscopy (SEM) with a JEOL JSM-840A (complemented with an EDAX EDS detector) at MNHN and a JEOL JSM6610-Lv (with an Oxford/SDD EDS detector) at the University of Toronto; X-ray maps were also obtained. Image analysis was performed using the JMicrovision software (www.jmicrovision.com): modes of petrographic components were determined by point counting (4000 points on each section) from BSE maps and chondrule effective diameters (that is, that of the equal area disks) were obtained from Mg X-ray maps. There is some ambiguity on the screening criteria for the chondrules; we will display results taking 0.1 mm as a minimum (apparent) diameter following King & King 1978), but given that the size distribution actually extends below it, making the threshold fairly arbitrary, we will also quote the area-weighted average (that is, the expected value of the enclosing chondrule radius of a random point in the chondrule population).

Minor and major element concentrations of documented chondrule phases, as well as isolated sulphide and matrix points were obtained with a Cameca SX-100 electron microprobe (EMP) at the Centre de Microanalyse de Paris VI (CAMPARIS). The beam current, diameter and accelerating voltage were 10 nA, ~5 μm and 15 kV, respectively. The following standards were used: diopside for Si, Ca; $MnTiO_3$ for Mn, Ti; pure iron for Fe; pure cobalt for Co; pure nickel for Ni; apatite for P; pyrite for S; sphalerite for Zn; $Cr_2O_3$ for Cr. Counting time was 10 s for Mg, Ca, Al, S, Cr, Mn, Ti, Ni, P, Co and Fe and 5 s for Si, Na and K. For metal and sulphide analyses, Cr metal and schreibersite were the standards for Cr and P. Typical detection limits were 0.03 wt% for $K_2O$, 0.06 wt% for $SO_2$, 0.08 wt% for $P_2O_5$, 0.04 wt% for $Na_2O$, 0.1 wt% for $TiO_2$.

### 2.2 Bulk chemical analysis

A ca 1 g sample was crushed using a boron carbide mortar and pestle into a homogeneous fine-grained powder in clean room conditions at the Institut Universitaire Européen de la Mer (IUEM), Plouzané. Major and some trace elements were determined by ICP-AES (inductively coupled plasma – absorption emission spectrometry) using a Horiba Jobin Yvon Ultima 2 spectrometer. In addition, trace element abundances were analyzed by ICP-SFMS (inductively coupled plasma-sector field mass spectrometry) using a Thermo Element 2 spectrometer. Details on the procedures can be found in Barrat et al. 2012).

### 2.3 Oxygen isotope analysis

Oxygen isotope measurements were carried out at the Stable Isotopes Laboratory of CEREGE. Molecular oxygen was extracted using the IR-laser fluorination technique (Alexandre et al. 2006; Suavet et al. 2010). The powdered samples (1.5 mg, 100-400 μm grain size) were heated with a 30 W $CO_2$ IR laser (Merchanteck) in the presence of $BrF_5$ (80 hPa for quartz standard, 150 hPa for NWA 5958). The released gas was purified through two cryogenic nitrogen traps and one heated KBr trap. Molecular oxygen was then 1) trapped for 10 min by adsorption in a microvolume filled with 13X molecular sieve and cooled at -196°C; 2) expanded at 100 °C and passed through a -114 °C slush for 5 min to refreeze possible interfering gases; 3) trapped again for 5 min in the microvolume cooled at -

196 °C; 4) expanded at 100°C directly to the sample bellow of the dual-inlet mass spectrometer (Thermo-Finnigan Delta Plus).

The isotopic compositions are expressed in standard δ-notation, relative to Vienna standard mean ocean water (VSMOW): $\delta^{18}O = (^{18}O/^{16}O)_{sample}/(^{18}O/^{16}O)_{VSMOW} - 1$ and $\delta^{17}O = (^{17}O/^{16}O)_{sample}/(^{17}O/^{16}O)_{VSMOW} - 1$ (both to be expressed in ‰). $\Delta'^{17}O$ is defined as $\ln(1+ \delta^{17}O) – 0.5247 \times \ln(1+ \delta^{18}O)$, also expressed in ‰ (Miller 2002).

Measured $\delta^{18}O$ and $\delta^{17}O$ values of the samples were corrected on a daily basis using the quartz laboratory standard (Boulangé 2008 50-100 $\mu$m) itself calibrated against the international standard NBS28 ($\delta^{18}O$ = 9.6±0.123‰, $\delta^{17}O$ = 5.026 ±0.075 ‰ and $\Delta'^{17}O$ = 0 ±0.026 ‰, $n$ = 23). Reproducibilities of the quartz laboratory standard $\delta^{18}O_{Boulangé}$ and $\delta^{17}O_{Boulangé}$ analyses (±1σ) are respectively 0.119 and 0.061 ‰ ($n$ = 63). The $\Delta'^{17}O_{Boulangé}$ value is -0.015±0.024 ‰ (1σ). $\delta^{18}O$, $\delta^{17}O$ and $\Delta'^{17}O$ values of quartz standards and reproducibilities (1σ) are similar ($n_{Boulangé}$ = 13, $n_{NBS28}$ = 8) when the refreezing step in the -114 °C slush (steps 2 and 3 described above) is included in the oxygen extraction procedure.

Three powdered 1.5 mg samples of NWA 5958 were prepared by gentle crushing of three separate bulk samples with mass ranging from 10 to 50 mg. These bulk samples were spatially separated by about 2 cm in the meteorite.

### 2.4 Magnetic measurements

The low field specific susceptibility and its evolution with temperature were measured using an Agico MFK1 apparatus equipped with a CS3 furnace and a CSL cryostat. Saturation magnetization was measured with a Princeton Micromag Vibrating Sample Magnetometer with a maximum applied field of 1 T. We measured the S ratio, defined as the isothermal remanent magnetization (IRM) obtained after applying a 3 T field and then a back field of -0.3 T normalized to the 3 T IRM. All magnetic measurements were performed at CEREGE.

### 2.5 Infrared spectroscopy

About 20 mg of NWA 5958 was dry grinded in an agate mortar. Out of this mass, 1.0 mg was weighed and mixed with 300 mg of commercial ultrapure potassium bromide powder (KBr). This mixture was then compressed to 400 bars in order to obtain a 13 mm diameter pellet of good optical quality. This whole procedure was reproduced twice to assess possible sample heterogeneity. Mid-infrared spectra were measured with a Brucker V70v spectrometer at the Institut de Planétologie et d'Astrophysique de Grenoble, following previous studies (Beck et al. 2014). Spectra were acquired at 2 cm$^{-1}$ spectral resolution in the 5000-400 cm$^{-1}$ range (2 to 25 μm). In order to investigate the hydrous mineralogy of NWA 5958 from the 3-μm region, the pellet was heated in an oven at 150°C and then at 300°C (each time for two hours) to remove weakly bound water molecules (from the sample itself as well as from the KBr).

## *3. Results*

### 3.1 Petrography and mineralogy

Global back-scattered electron (BSE) and X ray maps of NWA 5958 are displayed in Figure 1 and modal data are presented in Table 1. Individual EMPA data are listed in the Electronic Annex. Petrographic manifestations of terrestrial weathering are very minor, save for a few ~0.1 mm wide fractures filled with terrestrial minerals (quartz, feldspars, carbonates etc.) such as the subvertical one apparent in the middle of Fig. 1b; in particular little oxide veining is observed, with fracturing itself being very limited, consistent with the weathering grade W1 assigned by Bunch et al. (2011). Petrographically, NWA 5958 evokes overall a weakly altered CM chondrite, with 76 vol% matrix, 19 vol% chondrules, 2 vol% refractory inclusions and 4 vol% opaque material (the latter comprising about 1-2 vol% sulphides (from X-ray maps for S), <1 vol% metal and PCP) and an overall brecciated texture, if with various degrees of definition of the clasts (Bunch et al. 2011). A useful benchmark will thus be the Paris meteorite, the least altered CM chondrite hitherto described (Hewins et al. 2014), starting with the detailed description of the different components below.

The vast majority of the chondrules (Fig. 2) are porphyritic. While barred olivine (BO) textures also occur, no radial pyroxene one has been observed (although such textures represent 2 % of chondrules in the chondrule-wise similar CO chondrites; Jones 2012) possibly owing to the easier aqueous alteration of their slender enstatite laths (a possible example being depicted in Fig. 2e). The largest chondrules observed are 1 mm across. The average diameter of > 0.1 mm chondrules was 0.18 mm (n = 629, σ = 0.11 mm), 0.18 mm (n=342, σ = 0.11 mm) and 0.17 mm (n = 301, σ = 0.09 mm) for NWA 5958-1, 5 and 6, respectively, yielding a grand average of 0.18 mm (n=1272, σ = 0.1 mm). This is in the range found by King and King (1978) for CM and CO chondrites (see also Rubin & Wasson 1986; Rubin 1989; Friedrich et al. 2015). We also quote the (unthresholded) area-weighted averages as 0.32 mm (n = 4128, σ = 0.28 mm), 0.32 mm (n=1247, σ = 0.23 mm) and 0.26 mm (n = 1351, σ = 0.12 mm), respectively, hence a grand average of 0.3 mm. Chondrules are typically surrounded by fine-grained rims, sometimes multi-layered (Bunch et al. 2011).

Chemically, most chondrules are of type I (Fig. 2a-d), here defined as bearing ferromagnesian silicates with Fe/(Mg+Fe) < 10 mol% (regardless of texture). They generally display the characteristic mineralogical zoning of type I chondrules (Libourel et al. 2006), with olivine dominating in the inner regions whereas pyroxene dominates in the margin, with olivine chadacrysts often enclosed. While mafic silicates are quite pristine, mesostasis, unless entirely enclosed in the former, is generally altered, save for augite laths, in a fibrous or acicular Ca-depleted material (likely phyllosilicate); yet the small chondrule N5-24 retains fairly well-preserved mesostasis (Fig. 2d), suggestive of local heterogeneities in chondrule alteration (we note that one such chondrule was also reported in CM chondrite Elephant Moraine 96029 by Lindgren & Lee 2015), which meteorite also shares with NWA 5958 a lack of petrofabric (Lindgren et al. 2015)). While many chondrules essentially consist of single olivine grains (rimmed by enstatite; e.g. Fig. 2c), most intact chondrules have multilobate shapes (in particular Fig. 2a, b), as described for the CO3.0 chondrite Yamato 81020 by Rubin & Wasson 2005), and common for CO and CM chondrites. Since the lobes frequently display mineralogical zonings (e.g. presence of interior olivine) similar to typical "single" type I chondrules (see Fig. 1b, 2a, b), these appear best interpreted as adhering chondrules; i.e. the multilobate chondrules may be compound chondrules (e.g. Gooding & Keil 1981; Wasson et al. 1995; Akaki & Nakamura 2005; see also Jacquet et al. 2013, 2015). Then, given that the plane of sectioning may miss adhesions, it may well be that *all* the larger chondrules are compound (as also suggested by Alexander and Ebel 2012). Alternatively, the multilobate chondrules may be irregular objects not melted enough for surface tension to impose a spherical shape as argued by Rubin and Wasson (2005). The distribution of Fe/(Mg+Fe) is displayed in Fig. 6a for olivine and enstatite and minor elements are plotted against FeO in Fig. 6b. Most olivine grains have fayalite contents <2 mol% but the pyroxene distribution is slightly shifted to FeO-richer compositions, with a deficit of values below 1 mol% ferrosilite (as in Paris; Hewins et al. 2014), plausibly as a result of Fe diffusion near the chondrule margin, although a similar observation applies to Semarkona (Jones and Scott 1989) and may require pyroxene to record more oxidizing conditions than olivine during chondrule cooling (Jacquet et al. 2015). The minor element correlations are similar to those seen in CM and CO chondrites (Brearley and Jones 1998; Hewins et al. 2014).

About 1 out of 8 chondrules are of type II (Fig. 2f-h). They are generally olivine-dominated, with fayalite contents up to 63 mol%. Their Fe/Mn atomic ratio averages 102, similar to Paris (Hewins et al. 2014) and CO chondrites (Berlin et al. 2011). Relict forsterite grains with ferroan overgrowths are not uncommon. The altered mesostasis may contain, in addition to alteration products, ferroan augite grains, small olivine grains (e.g. Wasson & Rubin 2003) and chromite, as well as sulfides. The shape of the type II chondrules is rarely spherical, and many euhedral olivine crystals also occur isolated in the matrix (sometimes with remnant chondrule attached; e.g. Fig. 2h), possibly loosened from friable mesostasis (Richardson & McSween 1978).

Calcium-aluminum-rich inclusions (CAI; Fig. 3a, b) are generally more or less sintered aggregates of nodules of spinel rimmed by diopside (an aluminian variety, with 3.6 wt% $TiO_2$ and 15.4 wt% $Al_2O_3$ was encountered only in CAI N1-16, along with a 4 µm refractory metal nugget), with variable amounts of olivine and perovskite. Neither anorthite nor melilite were encountered, possibly as a result of alteration. Amoeboid olivine aggregates (AOA; Fig. 3c, d), outnumbering CAIs by a factor of ~2.5, constitute a continuum with CAIs. Indeed, subcircular CAI-like nodules of spinel and diopside may be found inside them, although such phases generally occur as more irregular patches in AOAs. While some AOAs are fairly irregular in shape and porous (e.g. Fig. 3d; some of the pores possibly attributable to the leaching out of refractory minerals; e.g. Krot et al. 2004), most are fairly compact—and strictly speaking hardly "aggregates" anymore—, with continuous olivine (and generally no enstatite; e.g. Fig. 3c), similar (but rare) objects having been described in Murchison (Krot et al. 2004).

Little Fe-Ni metal is preserved, except enclosed in mafic silicates, most of which being altered to a Fe-Ni-S-O material (comparable to that referred to as type I PCP in CM chondrites; (Bunch & Chang 1980)). The few measurements performed indicate solar compositions (Ni content around 5 wt%; Co/Ni = 0.05), similar to Paris data (Marrocchi et al. 2014a). Most sulphides are pyrrhotite grains, with pentlandite often at their periphery, and sometimes as lamellae (Fig. 5c). Calcium carbonates (e.g. Fig. 5a) represent <1 vol% of the meteorite (judging from Ca X-ray maps). We did not find the carbon grains reported in the matrix and chondrules by Bunch et al. (2011).

Inclusions comparable to the "Poorly Characterized (or "Crystallized") Phases" (PCP) common in CM chondrites (e.g. Fuchs et al. 1973; Bunch and Chiang 1980; Tomeoka & Buseck 1985) occur at a level of ~2 vol% (Fig. 4b). They are typically enriched in S (presumably carried by tochilinite) toward the margin and are very ferrroan in their cores, the nominal FeO content given by EMP analyses exceeding typically 60 wt% (see Electronic Annex), higher than Paris PCPs, possibly indicating the presence of amakinite ($(Fe,Mg)(OH)_2$) in addition to cronstedtite (Marrocchi et al. 2014a, b).

The unusual xenolith N1-6 (Fig. 4d) deserves a particular description. This fragment (0.4x0.3 $mm^2$) contains several chondrules and chondrule fragments, most apparently originally of type I but with various degrees of FeO diffusion from the exterior. A relatively coarse-grained (grain diameters of a few microns) matrix occurs interstitial to the chondrules with a distinct zone dominated by (crushed?) augite. A portion of the xenolith is rimmed by an intergrowth of sulphide and olivine similar to those described by Kojima et al. (2003) in Bishunpur chondrules (and as independent objects) and which they ascribed to impact mobilization—a postaccretionary interpretation certainly compatible with the present arrangement. At any rate, the texture of N1-6 suggests significant impact reworking but whether it derives from the same parent body as the host (as the chondrule size might suggest, along with preliminary oxygen isotope measurements (to be published elsewhere) indicating $\Delta'^{17}O$ = -6 ‰) or not cannot be ascertained yet.

Matrix (including fine-grained rims) makes up about three fourths of NWA 5958. We note that TEM observations by Stroud et al. (2014) revealed the presence of fine-grained nominally anhydrous silicates, sulphides, phyllosilicates (including cronstedtite), with only small amounts of amorphous material or O-anomalous presolar grains (from NanoSIMS mapping). A vacuole lined by fibrous material (stoichiometrically similar to $Fa_{70}$ olivine, but EMP analysis (Electronic Annex) yields low totals and could not possibly avoid overlap of grains to be reliable) was found in section NWA 5958-1 (Fig. 5d), reminiscent of those (amphibole/olivine) described by Dobrica & Brearley 2014) in the Tischitz H3.6 chondrite. The most striking feature of NWA 5958 matrix (especially visible in section NWA 5958-1) is the existence of irregularly shaped FeO-enriched zones (by a few wt%) lined by ~10 µm thick even more FeO-rich boundaries (Fig. 5). They are not clasts like the one shown on Fig. 2 of Bunch et al. (2011)—which could nonetheless originate from such a region—, and are also distinct from FeO-rich strata noticed by Bunch et al. (2011) in fine-grained rims around chondrules, although the zones in question sometimes surround only one chondrule (but may be connected to larger neighboring regions in the third dimension). To our knowledge, a similar structure has only been described in Miller Range (MIL) 07687, an ungrouped C3 meteorite (Brearley 2012), where these were interpreted as oxide-oxihydroxide-rich alteration fronts. In the case of NWA 5958, though, the degree of alteration of chondrules does not seem to differ on either side of the fronts.

| Section | Type I chondrules | Type II chondrules | AOA | CAI | Opaques | Matrix |
|---|---|---|---|---|---|---|
| NWA 5958-1 | 18.2 | 2.3 | 1.5 | 0.7 | 3.5 | 73.9 |
| NWA 5958-5 | 16.4 | 1.6 | 1.1 | 0.6 | 3 | 77.4 |
| NWA 5958-6 | 15.2 | 3.1 | 1.8 | 0.6 | 4.2 | 75.2 |
| Mean | 16.6 | 2.3 | 1.5 | 0.6 | 3.6 | 75.5 |

**Table 1**: Modes in NWA 5958 (vol%)

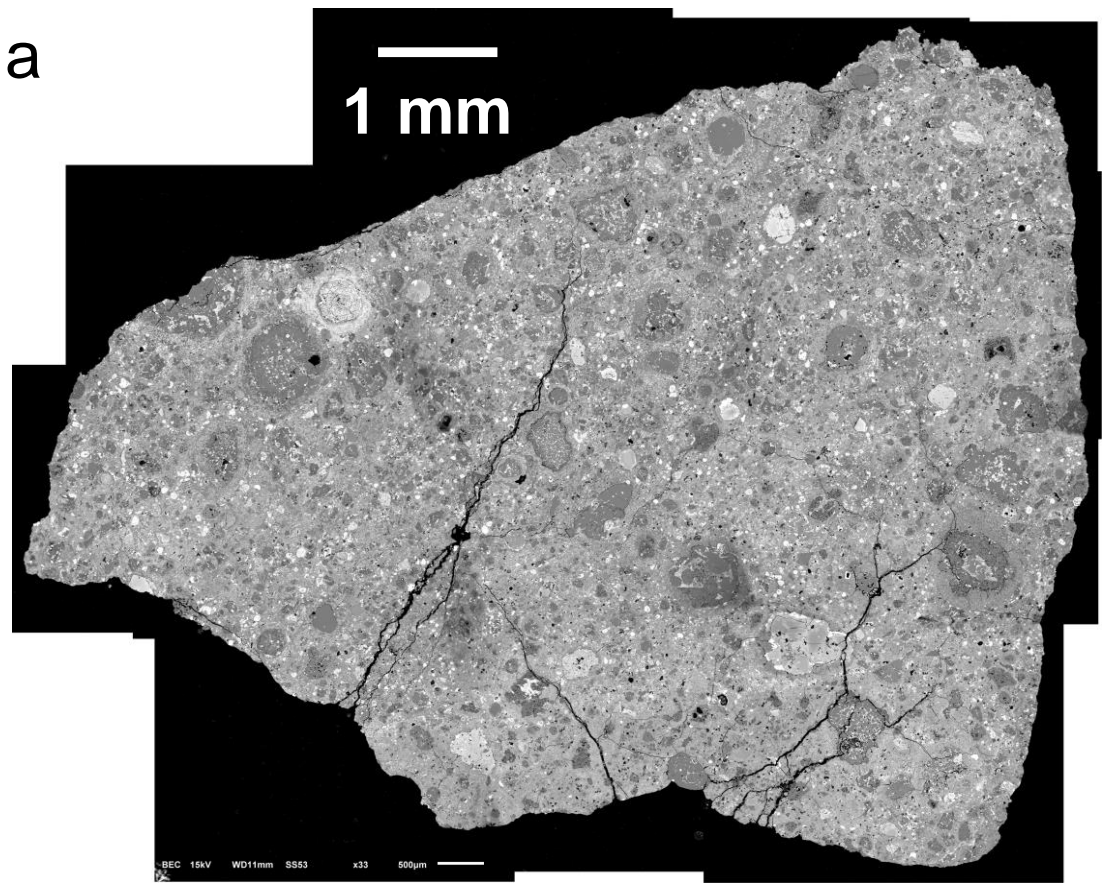

a

1 mm

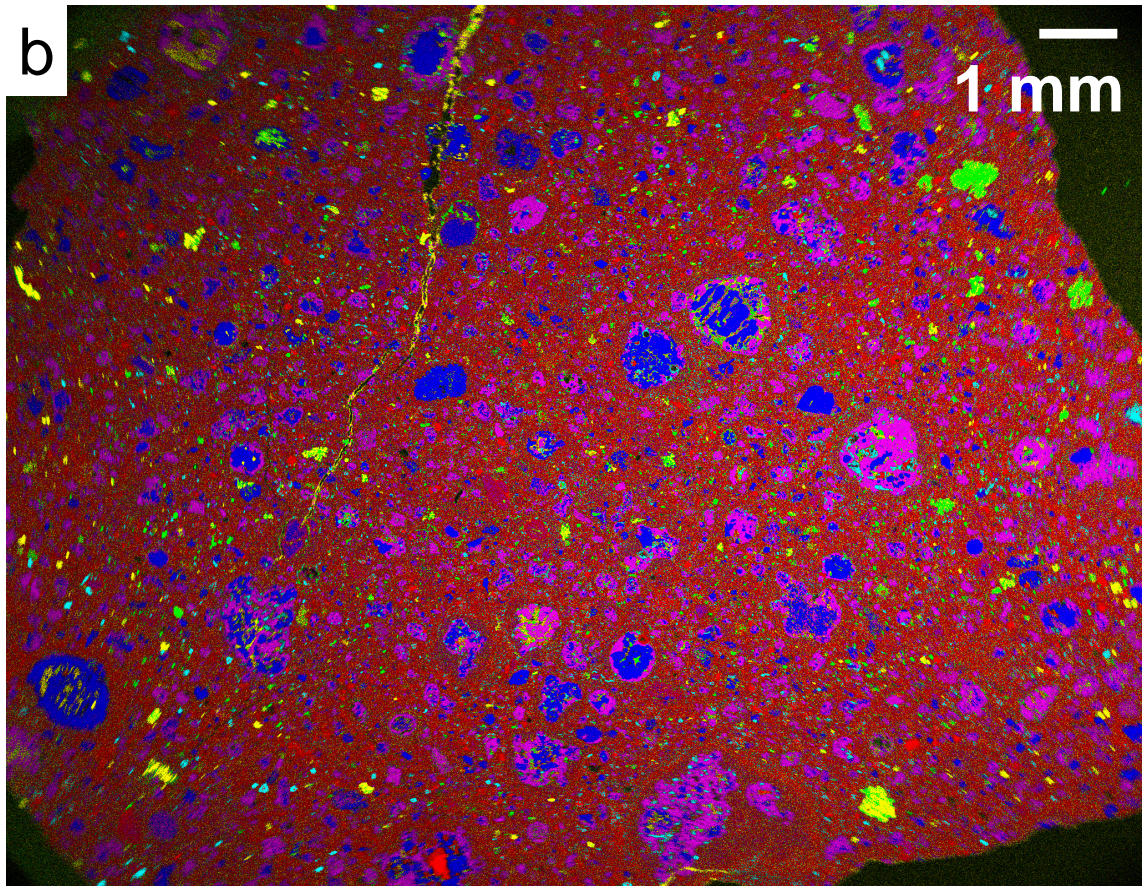

**Figure 1**: (a) Back-scattered electron map of section NWA 5958-6. (b) X-ray map of a portion of section NWA 5958-1. Color coding is Fe = red, Si = pink, S = cyan, Mg = blue, Ca = green, Al = yellow. Forsteritic olivine appears thus blue (while more ferroan one is more difficult to distinguish from the matrix) and enstatite pink in the type I chondrules ; the larger green/yellow inclusions are CAIs while smaller scattered green grains are mostly carbonate.

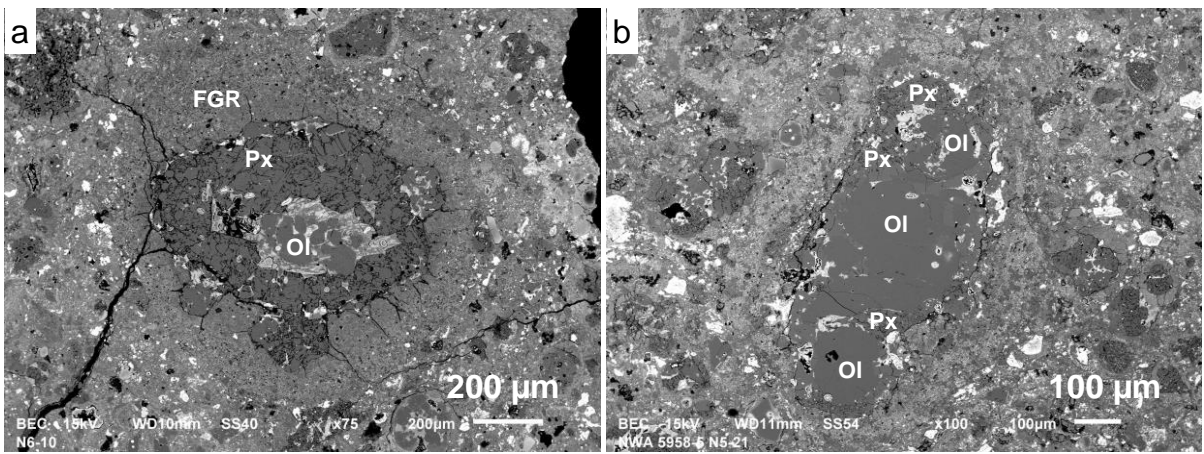

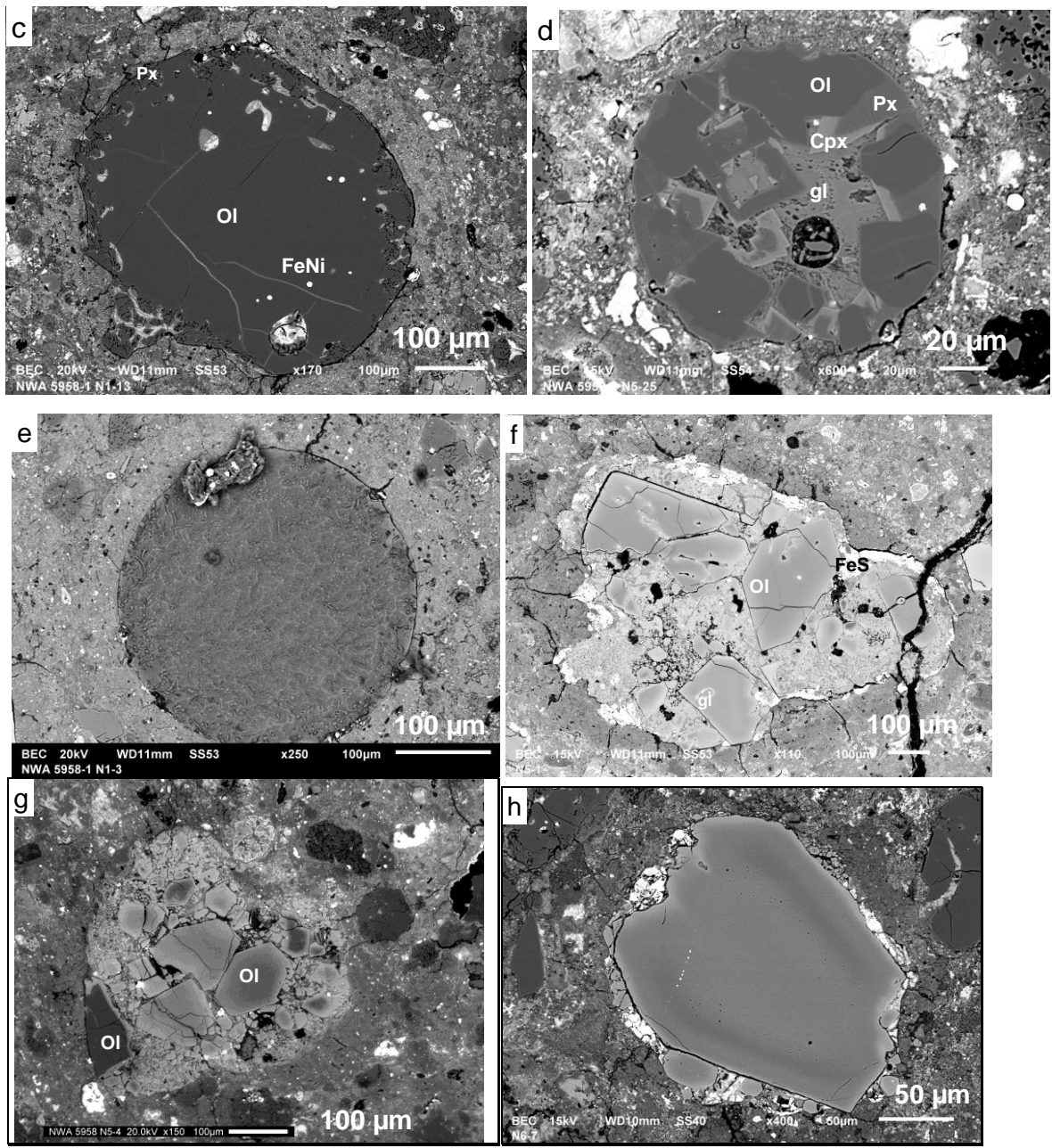

**Figure 2**: Back-scattered electron images of NWA 5958 chondrules. (a) Type I POP chondrule N6-10. Olivine and altered mesostasis dominate in the inner region whereas enstatite (with enclosed olivine chadacrysts) continuously dominates the outer region. At least three adhesions of mini-chondrules of similar mineralogical composition are visible on the surface of the chondrule, itself surrounded by a ~200 μm thick fine-grained rim (FGR). (b) Type I POP chondrule N5-21. This elongate chondrule apparently consists (in this plane of section) of essentially three aligned olivine grains, surrounded by enstatite. (c) Isolated ovoid forsterite grain, with sparse unaltered metal grains, partially altered vermicular mesostasis inclusions, and enstatite rim (with an apparent enstatite-mesostasis-rich adhesion in the lower left). (d) Type I POP chondrule. Despite its small ~120 μm diameter, it contains fairly well-preserved mesostasis at its center, with olivine and enstatite (sometimes hopper textured, overgrown by augite) dominating the outer region. (e) Nonporphyritic chondrule N1-3, now entirely altered, possibly originally of radial texture. (f) Type IIA porphyritic chondrule N5-1. Large normally zoned olivine crystals, with (sometimes intact) glass pockets, stand out in a mesostasis dominated by small olivine (and sparse chromite) grains. Troilite lines its multilobate surface. (g) Type IIA chondrule N5-4. A few forsteritic relict grains are evident, especially on the lower left, with a ferroan ~5-10 μm

thick overgrowth, and sharp contact with the matrix. (h) Isolated ferroan olivine crystal N6-7, presumably a fragment of a type II chondrule, as evidenced by still attached altered mesostasis, sulfides and smaller olivine phenocryst. The olivine presents an oscillatory zoning. Abbreviations: Ol = olivine, Px = low-Ca pyroxene, Cpx = Ca-rich pyroxene, gl = mesostasis, FGR = fine-grained rim.

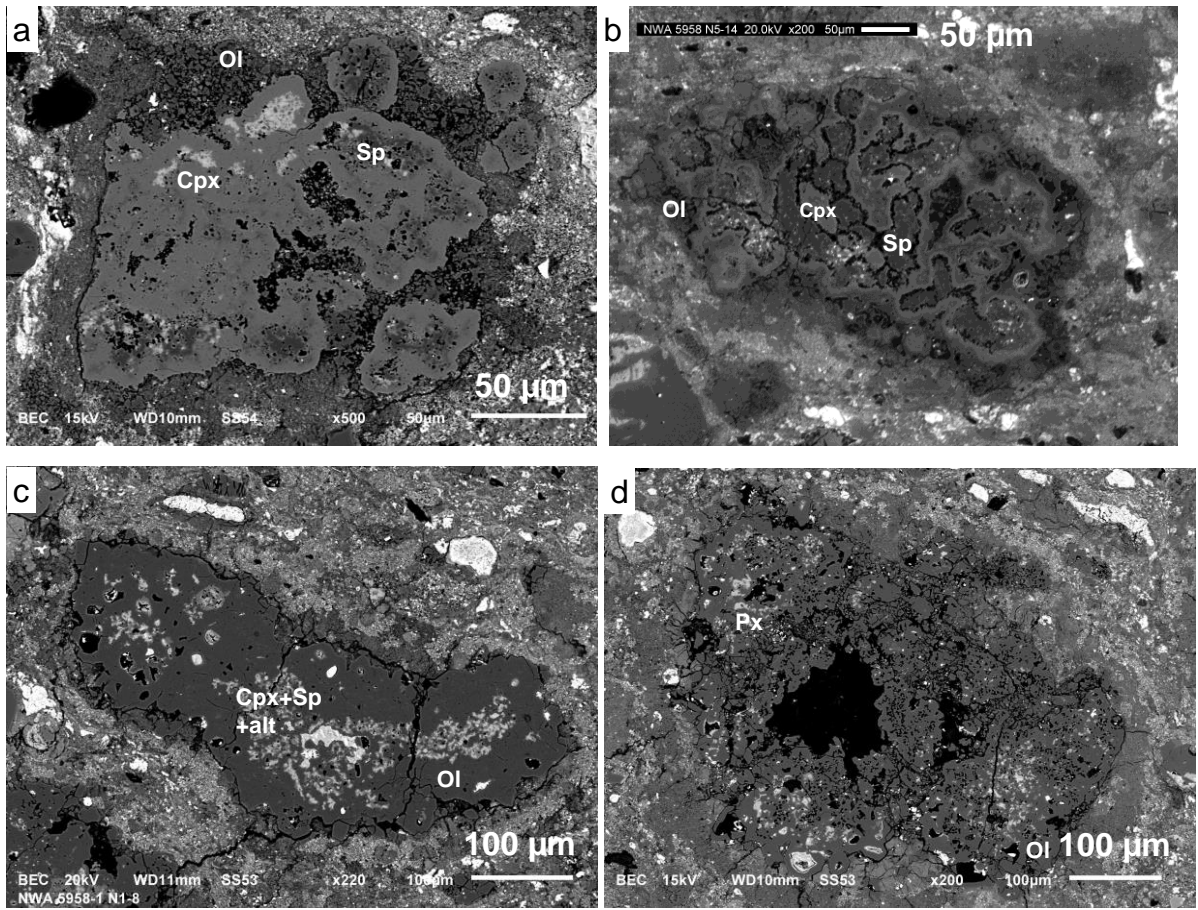

**Figure 3**: Back-scattered electron images of NWA 5958 refractory inclusions. (a) CAI N5-41, consisting of aggregated nodules of spinel grain clusters (unaltered to entirely altered) rimmed by diopside. Micron-size forsterite grains, grading into a fine-grained rim, partially surround the whole aggregate. (b) CAI N5-14 with continuous interconnected spinel nodules, peppered with perovskite, rimmed by diopside, the object as a whole being enveloped by continuous forsterite. (c) AOA N1-8, of compact texture, if elongate in shape, consisting of continuous forsteritic olivine, with patches of diopside, spinel and alteration products, a mineralogically zoned nodule being still reasonably preserved on the upper left. (d) AOA N1-19, much more irregular and porous than the former. In addition to the previous phases, minor patches of enstatite occur throughout the inclusion. Same abbreviations as Fig. 1 with Sp = spinel.

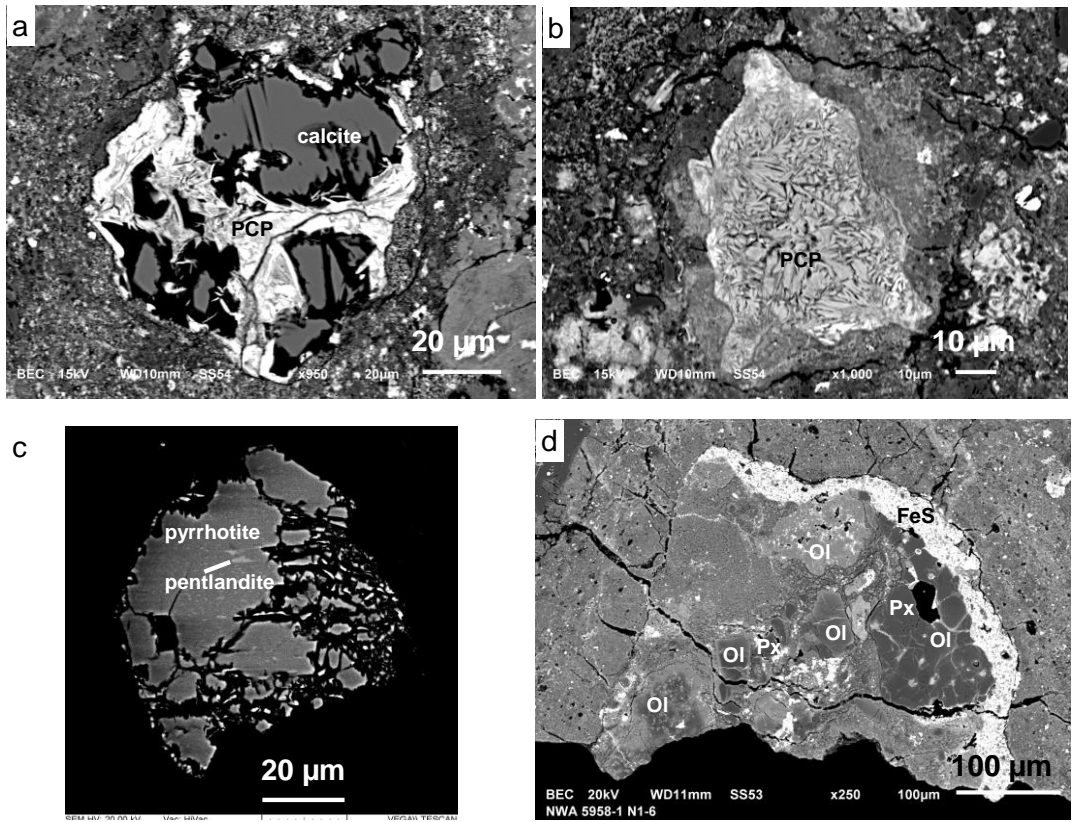

**Figure 4**: Back-scattered electron images of other inclusions in the NWA 5958 matrix. (a) Object N5-43 consisting of calcite intergrown with a fibrous PCP. (b) Fibrous PCP N1-22, with sulfur (tochilinite?) enrichment toward the margin. (c) Pyrrhotite grain, with pentlandite occurring as exsolution lamellae and at the margin. Contrast has been enhanced to distinguish between the two sulfides. (d) Xenolith comprised of several chondrules and chondrule fragments, with a relatively coarse-grained matrix (upper left being dominated by augite). The clast is partially rimmed by troilite (intergrown with silicates). Same abbreviations as Fig. 2.

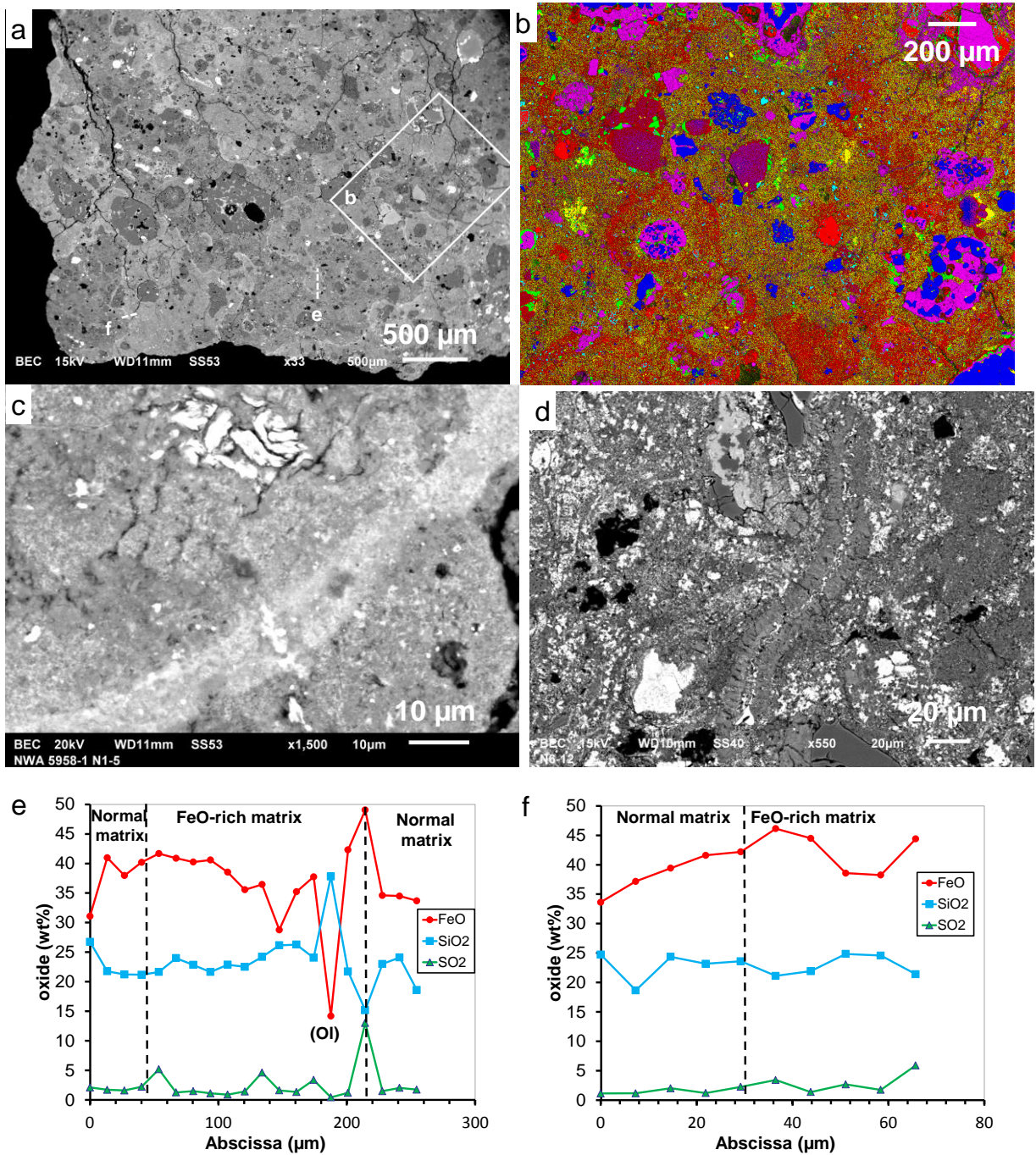

**Figure 5**: Images of the matrix of NWA 5958. (a) Global back-scattered image of an alteration front in section NWA 5958-1. Tracks denoted "e" and "f" are traverses shown in their respective panels below and zone "b" is zoomed in the following panel. (b) X-ray map of the (rotated) portion denoted "b" in panel a (same colors as Fig. 1b). Alteration fronts are FeO-enriched. (c) Close-up on the alteration front (near chondrule N1-5), with the FeO-enriched matrix on the upper side. (d) Elongate vacuole filled by fibrous material (near chondrule N6-12; analyses 6 and 7 in table S7, Electronic Annex). (e) and (f) Traverses whose tracks are depicted in panel a (e oriented from bottom to top and f from left to right in the orientation of panel a). Despite noise due to some inclusions (e.g. an olivine grain marked by "Ol"), one notes a transition from normal to FeO-rich matrix (and vice-versa, as approximately indicated) with a peak FeO enrichment near the boundary.

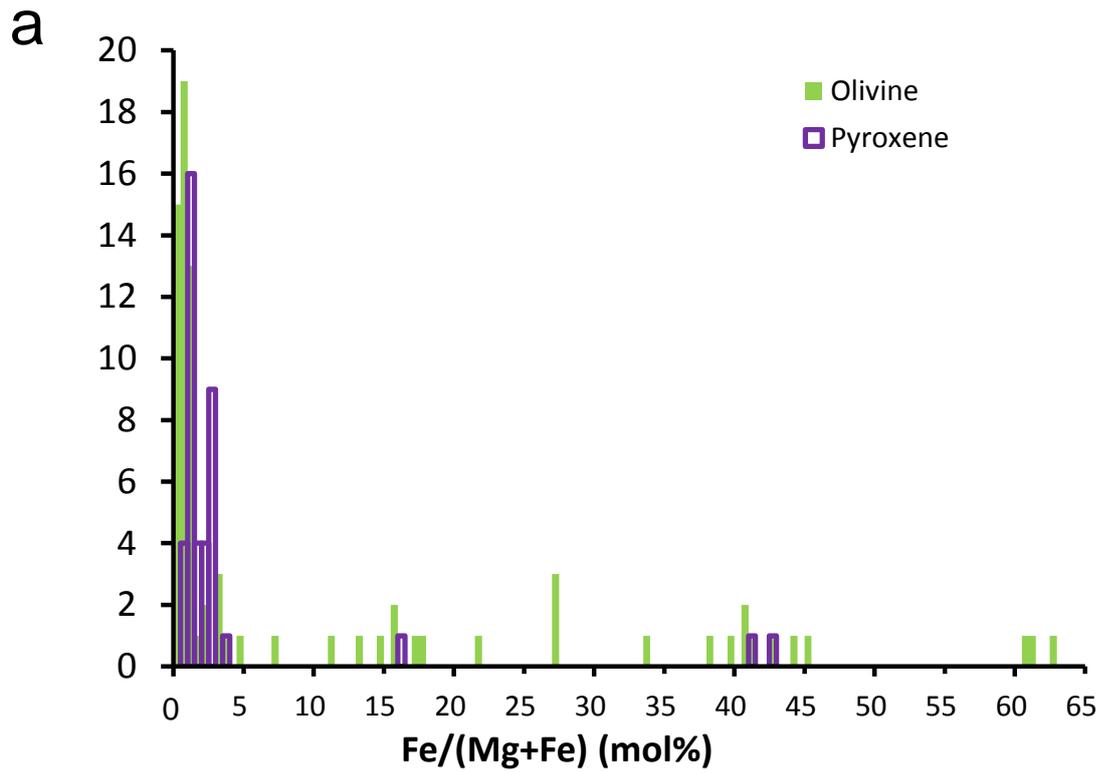

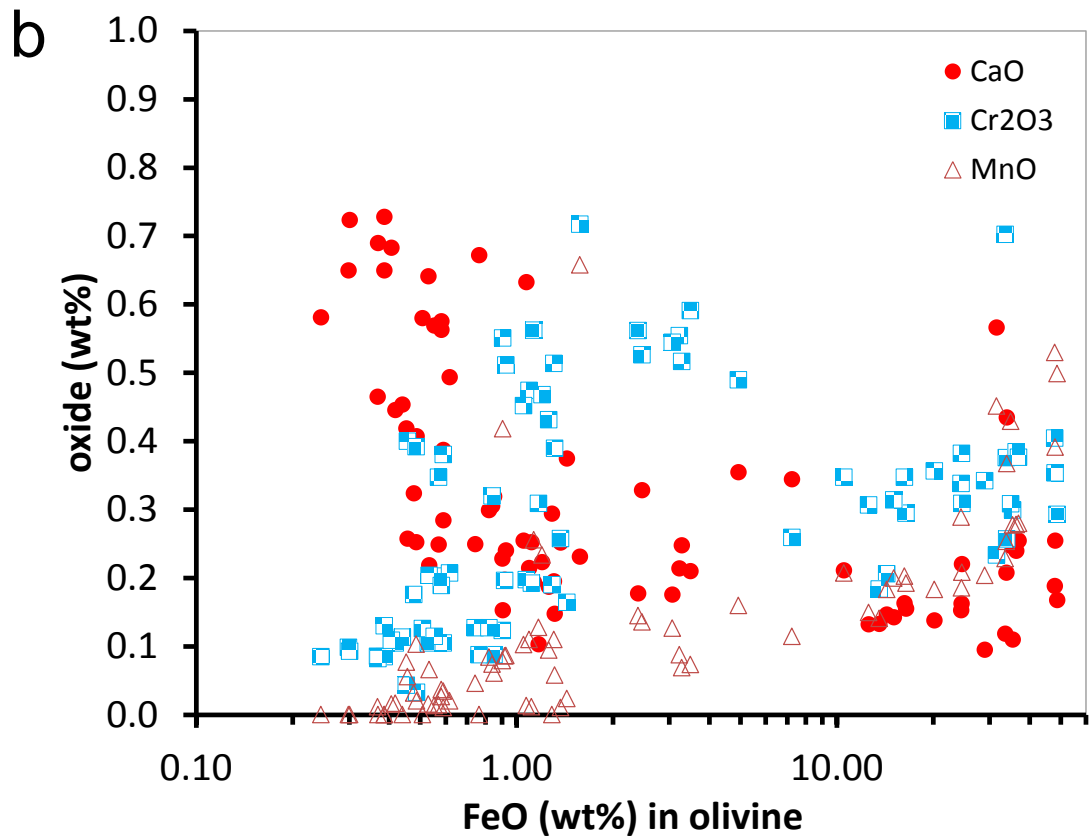

**Figure 6**: (a) Histogram of Fe/(Mg +Fe) in chondrule olivine and pyroxene. (b) Plot of CaO, $Cr_2O_3$ and MnO vs. FeO in chondrule olivine. X axis is here logarithmic to show type I and type II chondrules with similar relative resolution.

## 3.2 Bulk chemistry

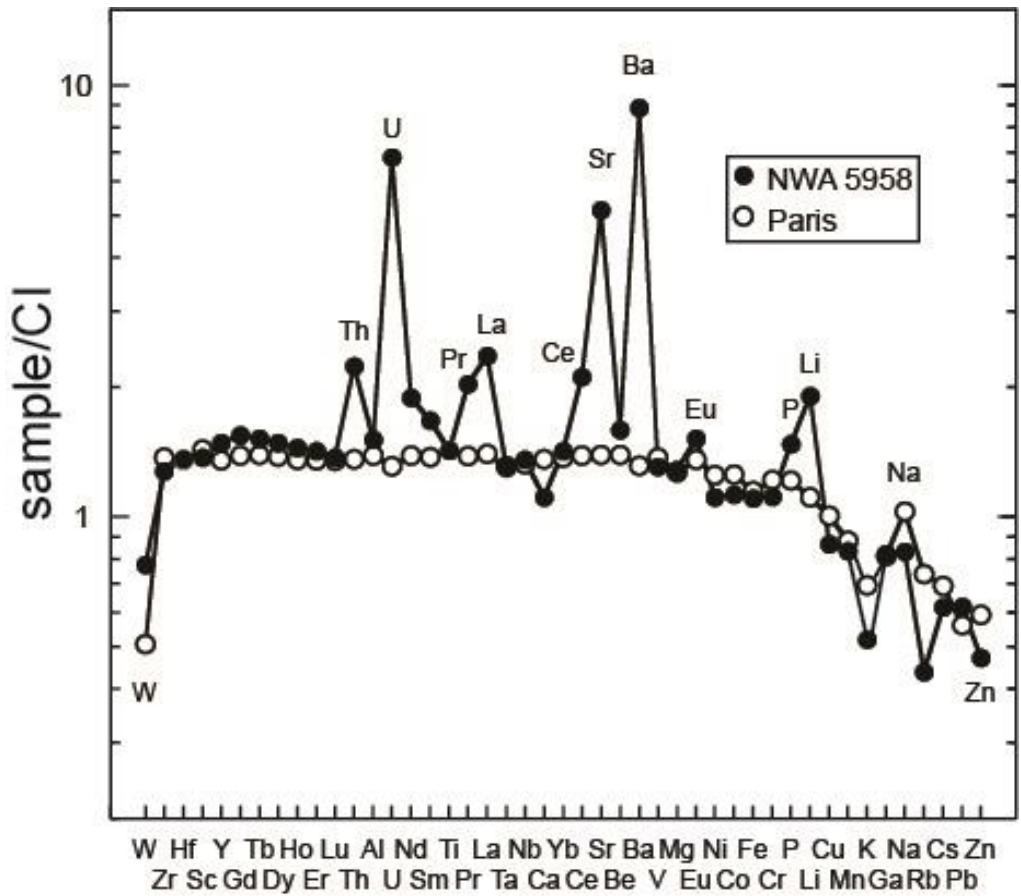

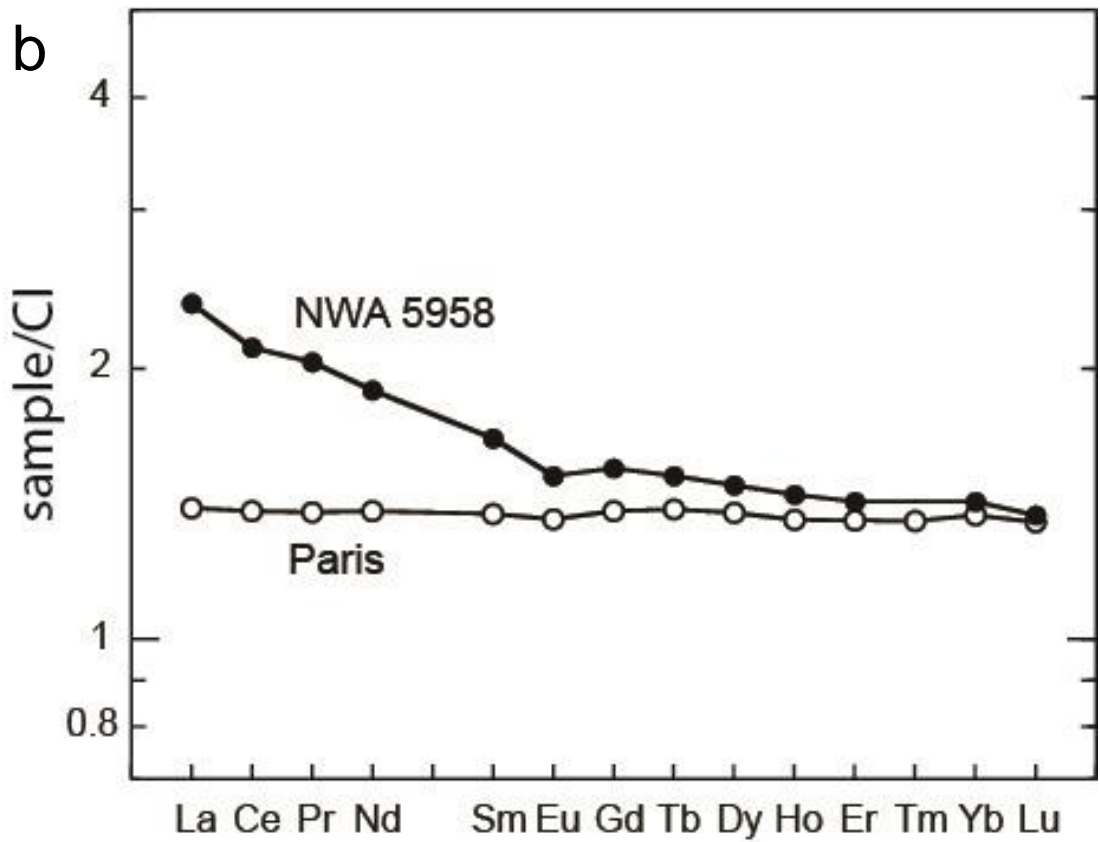

**Figure 7**: (a) CI-normalized concentrations of measured elements for NWA 5958 arranged in order of decreasing 50 % condensation temperature (Lodders 2003). The trend is similar to, if slightly steeper than Paris, although anomalies are superimposed by terrestrial weathering. (b) Rare Earth Element pattern of NWA 5958. Although the heavy REE roughly match Paris, the middle and light REE indicate a terrestrial contamination. In both panels, the CI values are taken from Barrat et al. 2012) and values for Paris (Hewins et al 2012) are plotted for comparison.

|  | NWA 5958 | CI | Paris |
|---|---|---|---|
| $SiO_2$ | 29.40 | 22.54 | 28.72 |
| $TiO_2$ | 0.11 | 0.07 | 0.1 |
| $Al_2O_3$ | 2.25 | 1.49 | 2.07 |
| $Fe_2O_3$ | 30.83 | 27.92 | 32.05 |
| MnO | 0.22 | 0.25 | 0.22 |
| MgO | 20.07 | 15.62 | 19.83 |
| CaO | 1.31 | 1.18 | 1.61 |
| $Na_2O$ | 0.54 | 0.65 | 0.67 |
| $K_2O$ | 0.09 | 0.07 | 0.04 |
| $P_2O_5$ | 0.34 | 0.23 | 0.28 |
| V | 74 | 52.4 | 72.1 |
| Cr | 2932 | 2627 | 3202 |
| Co | 586 | 519 | 651 |
| Ni | 12546 | 11300 | 14170 |
| Li | 2.75 | 1.44 | 1.6 |
| Be | 0.0360 | 0.0226 | 0.0315 |
| K | 293 | 550 | 380 |
| Sc | 8.04 | 5.85 | 8.38 |
| Mn | 1594 | 1910 |  |
| V | 68.5 | 52.4 | 72.1 |
| Cu | 110 | 127 | 128 |
| Zn | 143 | 303 | 180 |
| Ga | 7.80 | 9.48 | 7.71 |
| Rb | 1.02 | 2.33 | 1.72 |
| Sr | 39.81 | 7.73 | 10.79 |
| Y | 2.32 | 1.56 | 2.11 |
| Zr | 4.50 | 3.52 | 4.83 |
| Nb | 0.393 | 0.289 | 0.384 |
| Cs | 0.117 | 0.189 | 0.131 |
| Ba | 21.85 | 2.46 | 3.24 |
| La | 0.555 | 0.235 | 0.329 |
| Ce | 1.266 | 0.6 | 0.833 |
| Pr | 0.185 | 0.091 | 0.126 |
| Nd | 0.877 | 0.464 | 0.644 |
| Sm | 0.256 | 0.153 | 0.211 |
| Eu | 0.0891 | 0.0586 | 0.0797 |
| Gd | 0.319 | 0.206 | 0.286 |
| Tb | 0.057 | 0.0375 | 0.0523 |
| Dy | 0.377 | 0.254 | 0.351 |
| Ho | 0.082 | 0.0566 | 0.0769 |
| Er | 0.236 | 0.166 | 0.225 |
| Yb | 0.239 | 0.168 | 0.231 |
| Lu | 0.0338 | 0.0246 | 0.0332 |
| Hf | 0.146 | 0.107 | 0.144 |
| Ta | 0.0194 | 0.0148 | 0.0192 |
| W | 0.142 | 0.183 | 0.093 |
| Pb | 1.67 | 2.69 | 1.51 |
| Th | 0.0633 | 0.0283 | 0.0386 |
| U | 0.0525 | 0.0077 | 0.0101 |

**Table 2:** Bulk composition of NWA 5958, compared to CI chondrites (Barrat et al. 2012) and Paris (Hewins et al. 2014). Trace element abundances are given in µg/g except for oxides (ICP-AES), expressed in wt%.

The bulk chemical composition of NWA 5958 is reported in Table 2 and CI-normalized values are plotted in Figures 7a and 7b. NWA 5958 shows a depletion in moderately volatile element alike regular CMs as exemplified by Paris (Hewins et al., 2014), but with more scatter. Most of it likely reflects terrestrial weathering. Indeed, high Ba, Sr, U abundances (9, 5 and 7 x CI, respectively), with a high U/Th ratio (=0.83), and a light REE enrichment are all known fingerprints of hot desert weathering as discussed in a range of meteorite types (e.g. Barrat et al. 1999, 2003, 2010; Stelzner et al. 1999; Crozaz et al. 2003; Göpel et al. 2015). These contaminations are presumably mostly carried by the large fracture fillings (covering on the order of 1 % of the volume) rather than being present throughout the rock at sub-mm scale (as confirmed by preliminary LA-ICP-MS analyses of chondrite components, to be reported elsewhere).

These data stand in stark contrast to the preliminary solution and laser ablation ICP-MS ones in the abstract of Ash et al. (2011) which indicated CI abundances. We do not have any good explanation for this discrepancy. While Ash et al. (2011) report one order of magnitude higher enrichments in U, Ba, Sr than this study, a correlation between them and the moderately volatile elements would be difficult to understand and is not observed by these authors. Also, the petrographic description of NWA 5958 by Bunch et al. (2011) resembles our own and makes intrinsic sample heterogeneity, although real to some extent owing to the clastic nature of the meteorite, difficult to envision as an explanation. It is nonetheless possible that the laser beam of Ash et al. (2011) oversampled matrix material which is known to have a more CI-like composition than the bulk in carbonaceous chondrites (e.g. Bland et al. 2005; Zanda et al. 2012).

## 3.3 Oxygen isotopes

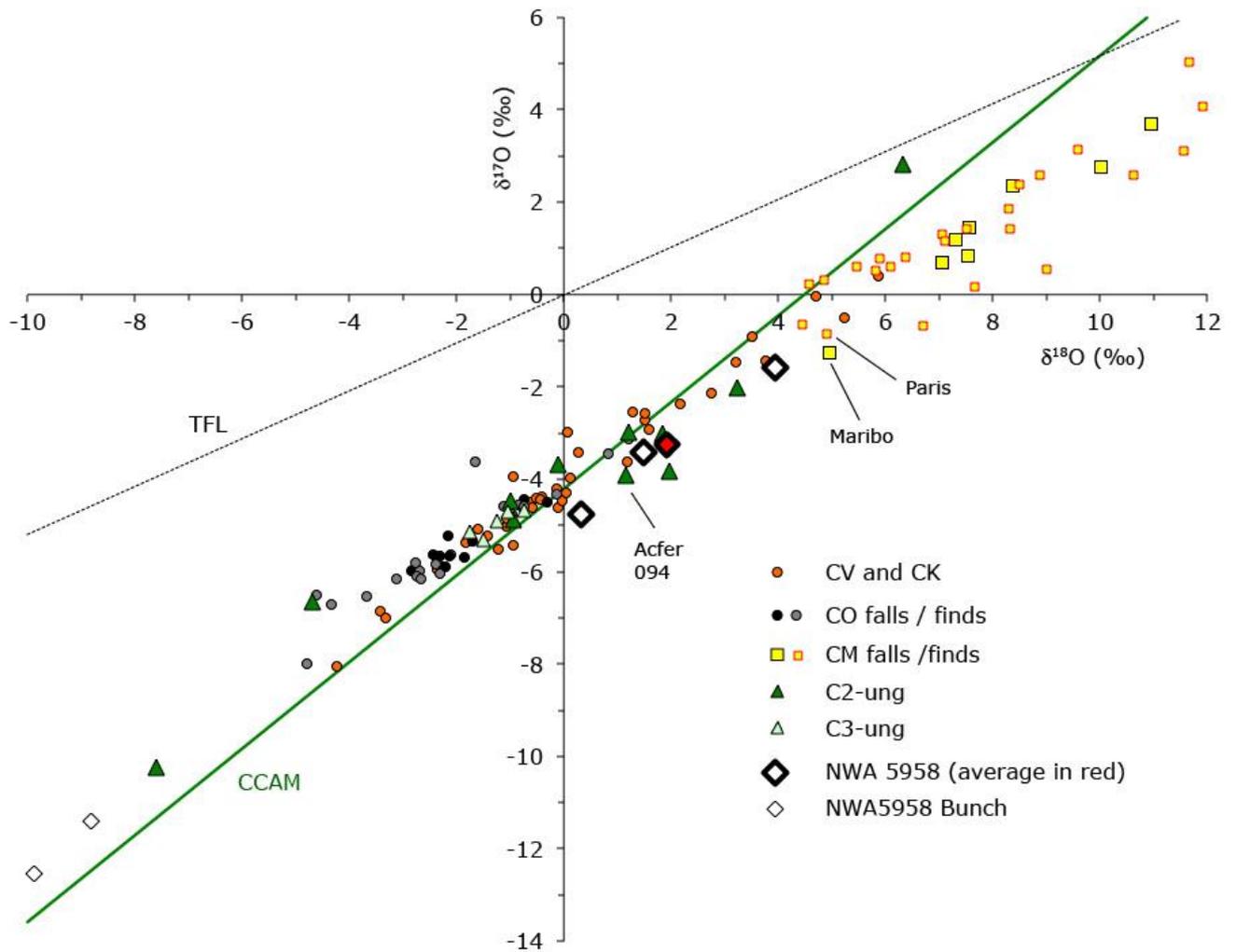

**Figure 8**: Three-isotope diagram with results from NWA 5958 compared with other carbonaceous chondrites (data from Clayton & Mayeda 1999; Greenwood & Franchi 2004; Haack et al. 2012; Hewins et al. 2014).

|      | δ$^{18}$O | δ$^{17}$O | Δ$^{'17}$O |
|------|-----------|-----------|------------|
| #1   | 0.33      | -4.75     | -4.94      |
| #2   | 3.93      | -1.58     | -3.64      |
| #3   | 1.49      | -3.42     | -4.20      |
| Mean | 1.92      | -3.25     | -4.26      |

**Table 3**: Oxygen isotope composition (in ‰) of NWA 5958.

Results of the bulk isotopic measurements are presented in Table 3 and plotted in Fig. 8. NWA 5958 lies near the CCAM, but with a more $^{16}$O-rich composition than CM chondrites, with a mean Δ$^{'17}$O = -4.3 ‰. NWA 5958 nevertheless lies close to values of several C2-UNG meteorites, like Acfer 094 (as opposed to C3-UNG which tend to overlap with the CO field).

We note the very different, albeit preliminary values of Bunch et al. (2011) who report Δ'$^{17}$O = -7 ‰ [note that the δ$^{17}$O and δ$^{18}$O are reversed in their abstract text, but not in their figure or ours]. Again, we do not have any good explanation, for it is difficult to envision how instrumental errors could move the composition along the CCAM. Preliminary in situ SIMS analyses of NWA 5958 chondrules (to be presented elsewhere) indicate that they all have Δ'$^{17}$O > -6 ‰ and refractory inclusions are in too small amounts (Bunch et al. 2011; this study) to satisfy the mass balance required by this extreme value, but, absent a specification of the size of the aliquots used by Bunch et al. (2011), it is conceivable that ($^{16}$O-rich) refractory inclusions were then oversampled. Whatever that may be, we note that our oxygen isotopic value would alone fit in the Δ'$^{17}$O - ε$^{54}$Cr trend of carbonaceous chondrites (Trinquier et al. 2007), taking the ε$^{54}$Cr = 1.0 ε determined by Göpel et al. 2015) and confirmed by Sanborn et al. (2015).

### 3.4 Magnetic properties

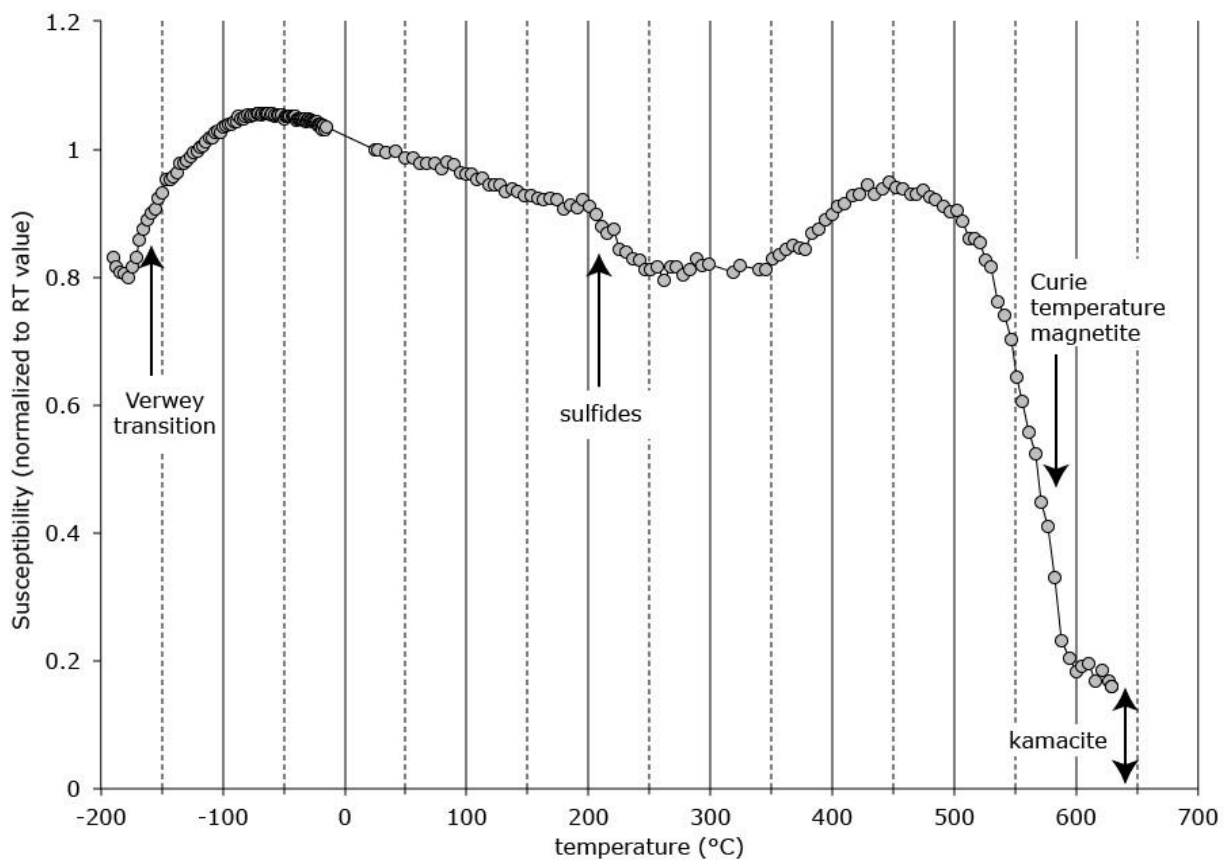

**Figure 9:** Evolution of low field magnetic susceptibility of NWA 5958 with temperature upon warming from -200 °C.

The magnetic properties are qualitatively similar to those of Paris (Hewins et al. 2014). Thermomagnetic experiments up to 630°C under argon atmosphere show a major and relatively smooth transition between 550 and 580°C (Figure 9). The computed transition temperature is 580°C, which is the Curie temperature of magnetite with no substitution. There is also a reversible transition at 210°C attributed to sulfides. Low-temperature magnetic measurements reveal a Verwey transition (Figure 9), confirming the presence of unsubstituted magnetite. The presence of ferromagnetic sulfides is also suggested by the S ratio value (-0.92) that indicates the presence of a high coercivity mineral. At 630°C, the magnetic susceptibility is 14 % of the initial one. This residual susceptibility is likely the signature of kamacite that has Curie temperatures above 700°C (the exact value depends on the Ni content). Kamacite is also evidenced by the curved shape of hysteresis cycles up to 1 T, a behavior typical of Fe-Ni grains with size >100 nm.

From the susceptibility drop across the transitions at 580°C (magnetite), and the residual susceptibility at 600°C (kamacite), the relative mass contents for magnetite and kamacite can be estimated to 2.9:1, assuming that all grains are multidomain and spherical. In view of the measured saturation magnetization (2.07 Am$^2$/kg), this indicates magnetite and metal contents of 1.2 and 0.42 wt%, respectively. The amount of ferromagnetic sulfides cannot be quantified by these magnetic measurements, but in view of the susceptibility drop around 210 °C, it is likely of the same order than the magnetite content (~ 1 wt%).

### 3.5 Infrared spectrum

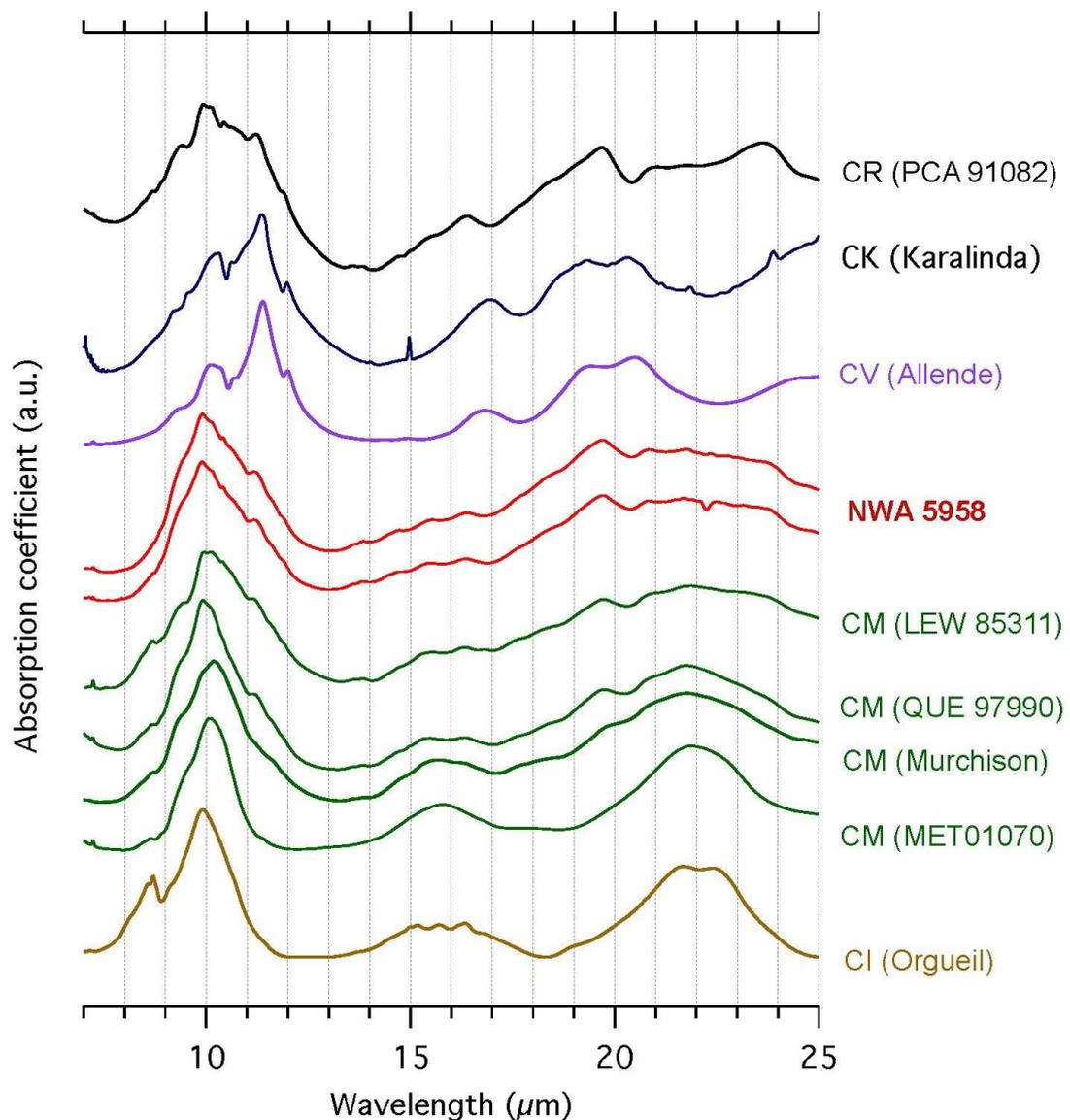

**Figure 10:** Transmission infrared spectrum of NWA 5958 in the 8-25 μm range (in arbitrary units). Absorption coefficients are also drawn for selected carbonaceous chondrites for comparison (Beck et al. 2014).

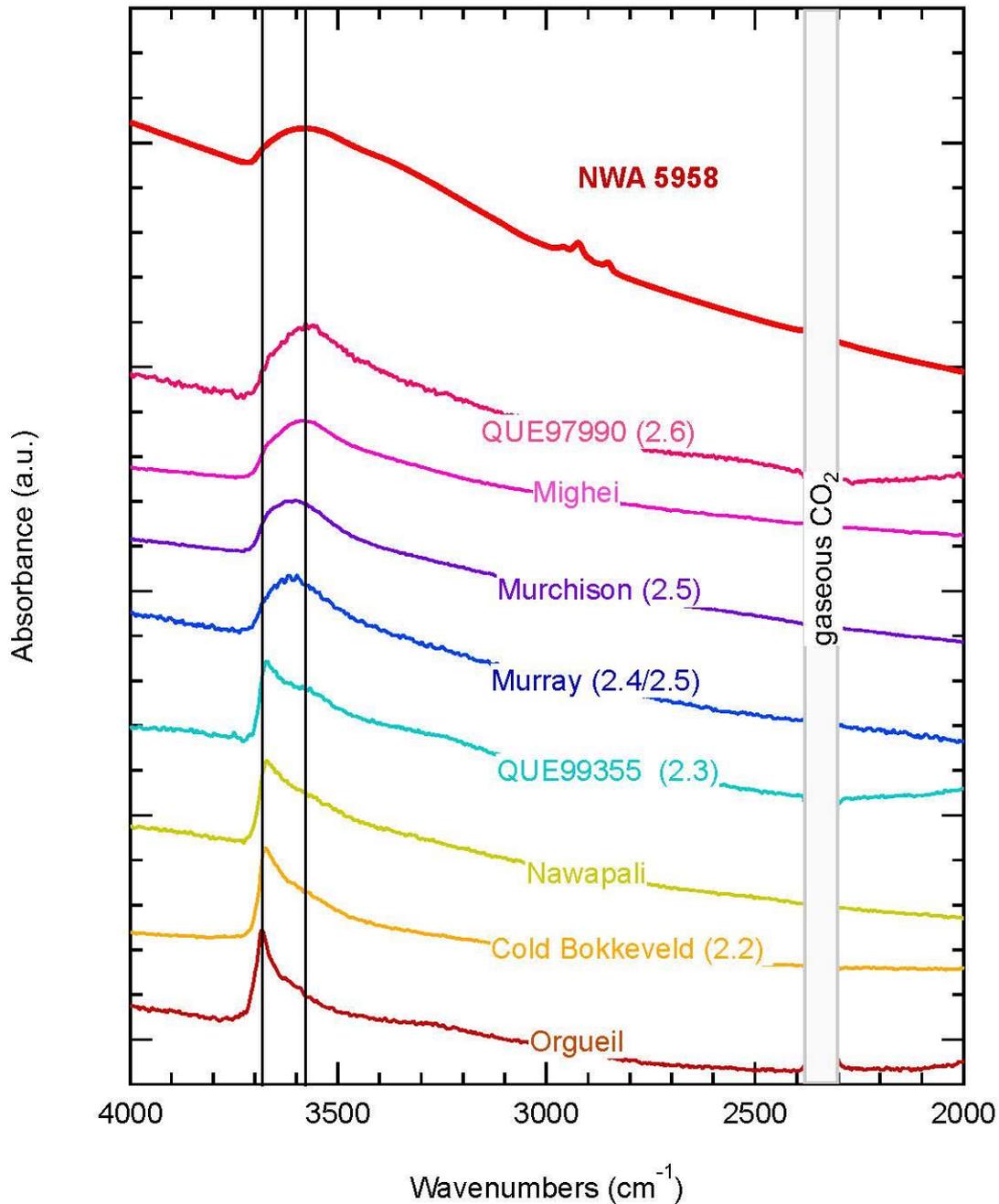

**Figure 11:** Transmission near infrared spectrum of NWA 5958 (in arbitrary units). Absorption coefficients are also drawn for selected carbonaceous chondrites for comparison (Beck et al. 2010).

The infrared spectrum of NWA 5959 in the 8-25 µm spectral range is presented in Figure 10. This region is dominated by $SiO_4$ stretching modes (8-12 µm) and bending modes at higher wavelengths. The two pellets prepared show quite similar infrared transmission spectra. A broad composite band with a maximum at around 10 µm is observed, that differs from typical spectra of CR, CK or CV chondrites (Fig. 10), but that resembles measurements obtained on CM chondrites. This feature is consistent with a mixture of phyllosilicates (with a maximum at around 9.9-10.0 µm) with forsteritic olivine (the shoulder at about 11.24 µm). In more details, the 10-µm region of NWA 5958 resembles weakly altered CM chondrites, the closest match being Lewis Cliff 85311 (which is incidentally a close match in terms of oxygen isotopic composition as well). The higher wavelength range of NWA 5858 infrared transmission spectra also resemble weakly altered CM chondrites (Fig. 10).

The presence of a significant fraction of hydrated minerals in the mineralogy of NWA 5958 is confirmed by the presence of an absorption feature in the 3 µm region (Fig. 11). This feature is relatively broad when compared to that measured on CM chondrites (Beck et al. 2010). The maximum of absorption is located at about 3580 cm$^{-1}$, also consistent with measurements performed on weakly altered CM chondrites. These characteristics of the 3 µm feature detected in NWA 5958 are suggesting that –OH group are present as Fe-OH rather than Mg-OH, possibly within tochilinites or Fe-bearing phyllosilicates (e.g. cronstedtite). The small spectral features around 2800-2900 cm$^{-1}$ (3.4-3.5 µm) are related to $CH_2$-$CH_3$ stretchings. While these features are present for chondritic insoluble organic matter (Orthous-Daunay et al. 2013), they cannot be unambiguously attributed to carbonaceous compounds from the meteorite itself, since they could also be produced by minor organic contamination in the KBr used.

## *4. Discussion*

### 4.1 Parent body processes

In this section, we assess the extent of secondary processes (metamorphism, aqueous alteration) experienced by NWA 5958, using, as a first approximation, indicators devised for CM and CO chondrites with which NWA 5958 has affinities. Since NWA 5958 is compositionally distinct and this study can only provide a rough overview, this of course is subject to revision by more focused future studies.

Grossman & Brearley 2005) identified the distribution of Cr content in type II chondrule olivine as a marker of incipient thermal metamorphism. In our data set for NWA 5958, we find a mean $Cr_2O_3$ content of 0.35 wt% and a standard deviation of 0.06 wt%, comparable to results for ALHA 77307 (0.38 and 0.07 wt%, respectively; Grossman and Brearley 2005), the most pristine CO chondrite, or Paris (0.37 and 0.08 wt%; Marrocchi et al. 2014a). Kimura et al. 2011) discussed the evolution of opaque phases during metamorphism of CM chondrites. The unsystematic presence of pentlandite lamellae in pyrrhotite and the solar composition of metal grains seem to place NWA 5958 in Kimura et al. (2011)'s category A, indicating metamorphic temperature below 300 °C. In comparison, Marrocchi et al. (2014a) ascribed Paris to category B (~300-750 °C).

NWA 5958 has undergone appreciable aqueous alteration, more consistent with a type 2, as favoured by Elmaleh et al. (2012) and Stroud et al. (2014), than a type 3 (Bunch et al. 2011). Rubin et al. 2007) proposed a classification scheme for progressive aqueous alteration of CM chondrites, in particular based on PCP. Following Rubin et al (2007)'s screening criteria, our PCP analyses (counting all Fe as FeO) yield an average "FeO"/$SiO_2$ = 5.0 (and S/$SiO_2$ = 0.05), higher than the CM2.7 chondrite Paris itself (4.1; Marrocchi et al. 2014a), which at face value would indicate lower levels of aqueous alteration (more limited to sulphides and metal hence the high abundance of oxidized iron in the alteration products). The infrared spectrum of NWA 5958 also resembles the least altered, albeit still hydrated, CM chondrites (section 3.5). Yet NWA 5958 displays levels of chondrule mesostasis replacement comparable to Paris and, most importantly, has preserved little metal (<~0.2 vol% from magnetic measurements, section 3.4) in comparison to the 2 vol% determined for Paris by Marrocchi et al. 2014a)—which would indicate a subtype 2.5 for NWA 5958 in the Rubin et al. (2007) scale. The style of alteration of NWA 5958 may have been quite distinct from regular CM chondrites, perhaps owing to more oxidizing conditions. Of course, we recall the notable peculiarity NWA 5958 presents with its alteration fronts in the matrix, which is only matched by MIL 07687 (Brearley 2012). Here however, the regions outside the alteration fronts are not pristine, so these fronts may reflect a second, limited, alteration episode following more extensive initial alteration. As decay of short-lived radionuclides such as $^{26}$Al may not have caused two different alteration episodes, the latter one may be due to an impact.

To summarize, NWA 5958 seem to have undergone little metamorphic heating, but significant, if moderate aqueous alteration, in at least two episodes. There is little point, here, to try to assign a precise petrographic type to NWA 5958, not only because metamorphism and aqueous alteration do not necessarily lend themselves to a one-dimensional scheme especially for near-pristine chondrites (Marrocchi et al. 2014a) but also because such classifications must *in fine* be tailored to specific chemical groups. In fact, as we shall now see, NWA 5958 may not neatly fit in an established group.

### 4.2 Compositional classification

From a petrographic standpoint, the major chemical group closest to NWA 5958, with e.g. its small chondrules, relatively abundant matrix, the presence of PCP, etc. is obviously the CM group. The bulk chemical composition of NWA 5958 (Fig. 7) is also roughly consistent with CM although it seems somewhat more volatile-depleted. For instance, the taxonomically useful Zn/Mn ratio of 0.076 is intermediate between CM and CO (Friedrich et al. 2002) and Ga/Ni, at 0.00052 lies at the lower end of the CM field (Scott & Newsom 1989), likewise trending toward COs. The oxygen isotopic composition (Fig. 8) provides a further, clear difference with known CM, with an enrichment in oxygen-16 (still less pronounced than CO), even relative to Paris which may have been less altered by putatively $^{16}$O-poor water (Clayton & Mayeda 1999). This overall intermediate position between CM and CO suggests that NWA 5958 (officially a C3.0-UNG) should be reclassified as an anomalous, CM-related chondrite ; or in the terminology of the Meteoritical Bulletin, a C2-UNG—which indeed populate this field in the oxygen three-isotope plot (Fig. 8)— if by "C2" we simply mean a (primarily) CM-related stone rather than a specification on the aqueous alteration (whose relatively unique style in NWA 5958 may be a further argument for non-grouping).

Of course, there is something artificial in these taxonomic considerations. How about alternatively considering NWA 5958 (along other C2-UNG) as part of an *extension* of the CM field (and speak of "anomalous CM")? It evidently depends on how one defines a chemical group. Surely, the ideal is to match a group to one parent body (as defined prior to any collisional disruption) but we do not know how variable samples from one given chondritic parent body can be. The problem is compounded by the likelihood that accretion in the early solar system produced a continuum of parent body compositions, as can be specifically entertained between CM and CO chondrites (Young et al. 1999; Hewins et al. 2014; Russell et al. 2014). It would even be conceivable that the CM group as currently recognized, which is the most populous among carbonaceous chondrites (and with which many micrometeorites, which are less biased towards the inner solar system, bear affinities; Gounelle et al. 2005), is itself composite in origin, with perhaps some differences in primary compositions (e.g. oxygen isotopic; Clayton and Mayeda 1999) obliterated by aqueous alteration. Yet the present-day spectrum of chondrite group (14 in number) compositions is discrete, presumably as a result of the depletion of the asteroid belt mass by 1-3 orders of magnitude early in solar system history (Morbidelli et al. 2009; Weidenschilling 2011) and selectivity of meteorite delivery due to the position of orbital resonances, which allow us to sample only a minority of the initial planetesimals. Specifically for the CM group, its relative chemical homogeneity (Brearley 2003) and systematically young cosmic ray exposure ages (although, as a property shared by CI chondrites, perhaps simply a consequence of their fragility (Eugster et al. 2006); still, a cosmic ray exposure age would be worth determining for NWA 5958 and other C2-UNG) speak in favour of a single parent body for the majority of its members (possibly subject to tidal disruption at low velocity relative to the Earth according to Morbidelli et al. (2006)).

Then, could NWA 5958 originate from the CM parent body? Although, as said, the compositional variability of an asteroid is unknown, a tentative criterion might be derived from the presumption that asteroid accretion had to be rapid compared to protoplanetary disk evolution timescales (except in the case of layered accretion, this and other scenarios being discussed by Johansen et al. 2015)) so that throughout its extent its components have to be drawn from the same reservoir. Then primary compositional variations should reflect modal variations in petrographic components, e.g. owing to size sorting during accretion or subsequent reworking in the parent body (as well as statistical fluctuations depending on sample size). Here, NWA 5958 tends to be richer in matrix (76 vol%) than most CM chondrites (e.g. Zanda et al. 2006), and in particular Paris (63 vol% matrix according to Marrocchi et al. 2014a). Yet, it tends to be somewhat more volatile-depleted than them although CM matrix is generally close to CI chondritic composition (closer at least than the bulk composition; Bland et al. 2005; Zanda et al. 2012). Also, CM matrix tends to be depleted in oxygen-16 relative to the bulk (e.g. Clayton 2003), and while this may be affected by aqueous alteration (Clayton and Mayeda 1999), measurements in Acfer 094 suggest that some of this depletion is a primary feature (Sakamoto et al. 2007). Yet, NWA 5958 is *enriched* in $^{16}$O relative to regular CM chondrites, Paris included, in spite of having apparently undergone no less aqueous alteration than Paris. So it would appear that NWA 5958 originates from a distinct, if compositionally (and thus likely spatio-temporally) close, parent body than most CM.

*4.2.1 Implications for the two-component model?*

  Before closing this discussion, it may be noted that, prima facies, the latter considerations also bring some tension between NWA 5958 and the two-component model (Anders 1964; Zanda et al. 2006), where compositional variations among chondrites are ascribed to mixing between the same petrographic components *disk-wide* (and not merely within one local reservoir). Specifically, the two-component model envisions a CI component identified with the matrix and a high-temperature (HT) one, such that the concentration C of any element X in a chondrite is given by:

$C_X = (1-f_{CI})C_{HT, X} + f_{CI}C_{CI, X}$

with $f_{CI}$ the fraction of the CI chondritic component. For sufficiently volatile elements, the contribution of the high-temperature would be negligible so that $C_X \approx f_{CI}C_{CI}$, that is the CI-normalized abundance would plateau at the CI contribution fraction (or at least be an upper bound for it). In NWA 5958, the moderately volatile elements seem to plateau at ~0.5 x CI (Fig. 7a), making it anyway mathematically impossible for the matrix (76 vol% of the meteorite) to be CI chondritic—a classical assumption of the two-component model (Anders 1964; Zanda et al. 2006). This discrepancy was actually already noted as a general feature of carbonaceous chondrites by Zanda et al. (2012) who argued that part of the matrix may be due to alteration and/or the presence of volatile-depleted dust cogenetic with chondrules (e.g. originating from fine-grained accretionary rims; Metzler et al. 1992). The latter case may be consistent with evidence for matrix/chondrule complementarity for some elemental ratios (Hezel and Palme 2010; Palme et al. 2015), if confirmed (but see Zanda et al. 2012): Indeed, as illustrated in Fig. 12, mixing with a CI chondritic component (here about half of the meteorite) chondrules *and* dust complementary to the latter would still globally yield solar values for the elemental ratios considered, even though the matrix would be composite in origin (see also Jacquet 2014). In that sense, complementarity would be *consistent* with the *average* provenances of chondrule and matrix being different (as could be indicated by isotopic evidence, e.g. Trinquier et al. 2007, 2009), with no further assumption than a variation of the chondrule production rates across different reservoirs, and tight solid/gas coupling (Jacquet et al. 2012). This does not, of course, mean that matrix/chondrule complementarity and the two-component models are equivalent or predictively trivial. For matrix/chondrule complementarity, on the one hand, still needs to be quantified in an homogeneous manner (see Goldberg et al. 2015; Zanda et al. 2012), and, on the other hand, it has yet to be seen whether the bulk chemical and isotopic compositions of carbonaceous chondrites can be suitably modelled as a two-endmember mixing (Zanda et al. 2012), even though the said endmembers do not have simple petrographic manifestations. At any rate, the non-CI bulk composition reported here for NWA 5958 removes the most serious objection this meteorite would have presented to two-component models based on earlier preliminary data (Ash et al. 2011).

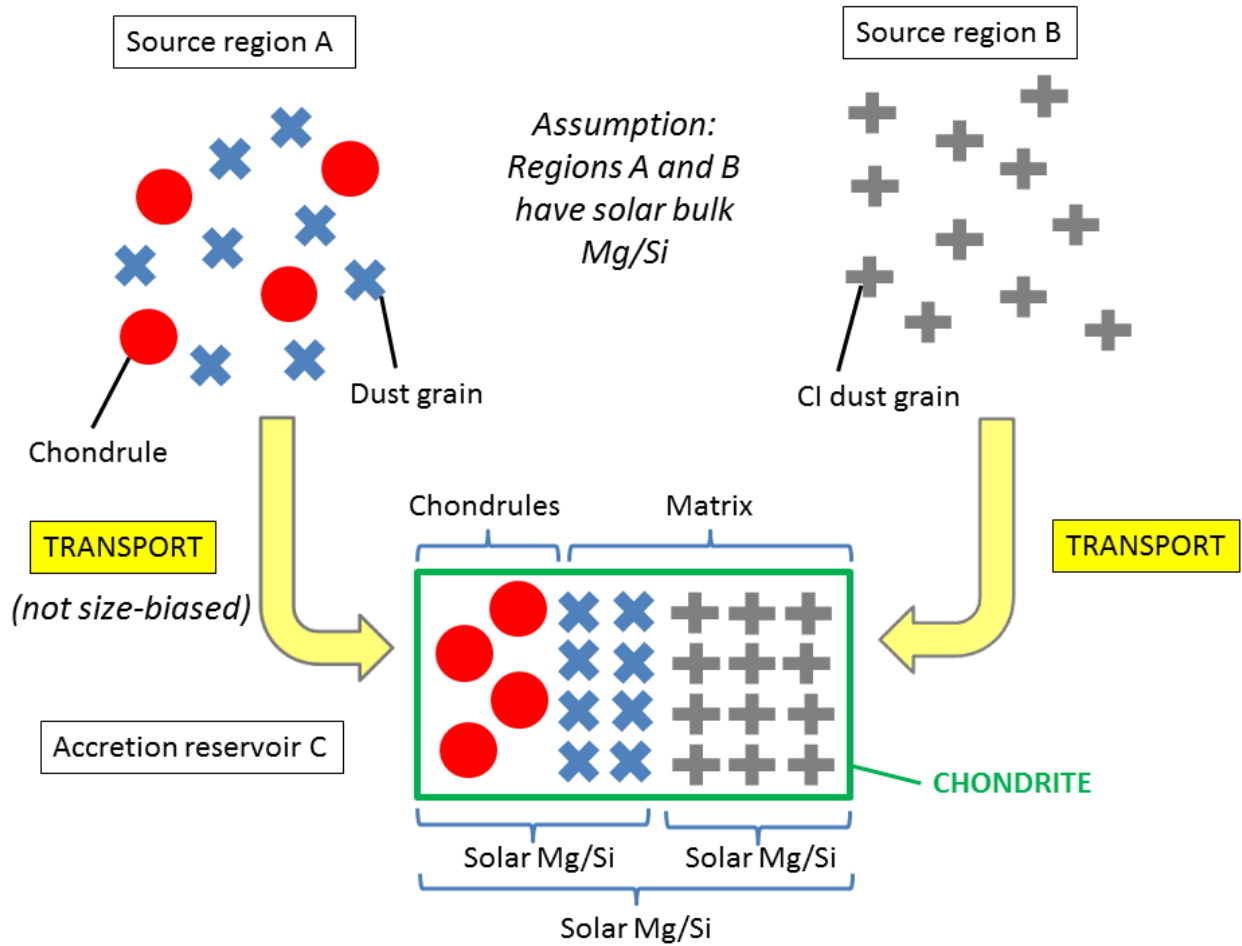

**Figure 12:** Sketch of a mixing between chondrite components arising from two source regions with a solar Mg/Si ratio, with region A having witnessed chondrule formation unlike region B. If transport does not distinguish between different sizes, the final product will also have a solar Mg/Si ratio and exhibit chondrule/matrix complementarity even though the sources are diverse. The reasoning can be generalized to an arbitrary number of sources and chondrule production efficiencies in these as well as any elemental ratio for which a solar initial value in these sources can be assumed.

## *5. Conclusion*

We have studied the Northwest Africa 5958 meteorite through a variety of techniques: scanning electron microscopy, electron microprobe, ICP-AES/ICP-SFMS, IR laser-assisted fluorination oxygen isotopic analyses, magnetic measurements and infrared spectroscopy. NWA 5958 is petrographically and chemically similar to CM chondrites, with somewhat more matrix than Paris. There is little evidence for thermal metamorphism and aqueous alteration seems to have been fairly weak as evidenced by the relatively low hydratation indicated by infrared spectra. Yet metal is extensively oxidized, not to mention the nearly ubiquitous alteration of chondrule mesostases, so it is certainly not a type 3.0 meteorite. An interesting feature is the presence of alteration fronts, not unlike MIL 07687, which in this case may reflect an overprint from a second, limited, possibly impact-induced alteration episode.

NWA 5958 is more $^{16}$O-rich than most known CM chondrites which, along with compositional arguments on the matrix/chondrule ratio, is interpreted to indicate a separate parent body of origin. We suggest that NWA 5958 be reclassified as an ungrouped, CM-related carbonaceous chondrite, sampling another parent body intermediate between CM and CO chondrite.

While NWA 5958 does not have a bulk CI composition, its matrix (as in other carbonaceous chondrites) cannot either. This does not exclude that *part* of the matrix is inherited from a CI composition component, and we have shown that this would *not* be incompatible with a picture where chondrules and (part of the) dust were cogenetic (complementary).

*Acknowledgments*: We thank Maximilien Verdier for fruitful discussions on CM chondrites. We are also grateful to Dr Alan Rubin and Dr Makoto Kimura for their careful reviews of the manuscript.